\documentclass[
  ,draft            
  ]
  {aipproc}

\layoutstyle{6x9}

\begin{document}

\title{Do Skyrme forces that fit nuclear matter work well in finite nuclei?}


\author{P. D. Stevenson}{
  address={Department of Physics, University of Surrey, Guildford, Surrey, GU2 7XH, United Kingdom}
}

\author{P. M. Goddard}{
  address={Department of Physics, University of Surrey, Guildford, Surrey, GU2 7XH, United Kingdom}
}

\author{J. R. Stone}{
  address={Oxford Physics, University of Oxford, Parks Road, Oxford, OX1 3PU, United Kingdom}
,altaddress={Department of Physics and Astronomy, University of Tennessee, Knoxville, Tennessee 37996, USA}
}

\author{M. Dutra}{
  address={Departamento de F\'isica, Instituto Tecnol\'ogico de Aeron\'autica, CTA, S\~ao Jos\'e dos Campos, 12228-900, SP, Brazil}
}

\begin{abstract}
A shortlist of Skyrme force parameterizations, recently found to have passed a series of constraints relating to nuclear matter properties is analyzed for their ability to reproduce data in finite nuclei.  We analyse binding energies, isotope shifts and fission barriers.  We find that the subset of forces have no common ability to reproduce (or otherwise) properties of finite nuclei, despite passing the extensive range of nuclear matter constraints. 
\end{abstract}

\maketitle


\section{Introduction}
A recent essentially exhaustive study of Skyrme parameterizations for their ability to satisfy experimental constraints of nuclear matter properties produced a shortlist of satisfactory parameter sets \cite{Dut12}.  No attention was paid in that study to the ability of the parameterizations to fit properties of finite nuclei.  In this contributions we analyse the shortlist of sets for their ability to reproduce various properties of finite nuclei.  We proceed by summarizing the nuclear matter constraints, then by discussing the properties the satisfactory parameter sets, including details related solely to the finite nuclear character, followed by a description and analysis of their behavior in finite nuclei, and conclude with a discussion of the results.

\section{The Skyrme Force}
The Skyrme interaction dates to the 1950s, when Skyrme first postulated his form of the nuclear effective interaction, based on the idea that it should be short-ranged, with a truncated Taylor expansion in powers of the relative momentum of the interacting particles \cite{skyrme}.  Rather than describe the Skyrme force in full detail, we refer the reader to recent reviews \cite{stone-reinhard,bender-review,example-skyrme} and list only what is necessary here.  As well as its functional form, the Skyrme force is defined by around ten (depending on the variant) parameters which must be fitted.  Fitting these parameters can be viewed as a minimization problem, where the fitness function has ten variables, and can be defined in various ways, most typically by a chi-squared function of calculated values of experimentally-known data.  Each Skyrme parameterization is generated by a combination of the fitness function, the minimization algorithm and the starting set of parameters used to initiate the algorithm.  The complexity of the multi-dimensional function and the freedom of choice of data to fit to has led to a large number of parameterizations.  In the next section, a summary of the nuclear matter constraints \cite{Dut12} is given, along with some details about the fitting procedures of the ``good'' forces is given.

\section{The Constraints and the Parameter Sets}
The recent set of nuclear matter constraints \cite{Dut12} used empirically-derived constraints on the nuclear incompressibility, the skewness coefficient, the pressure of both symmetric matter and pure neutron matter, the equation of state of neutron matter, the symmetry energy and its derivative at saturation density, and the half-saturation-density symmetry energy and the volume isospin incompressibility.  These macroscopic constraints gave an initial shortlist, which was further reduced by some microscopic constraints (e.g. Landau parameters) and observed neutron star properties to produce a shortlist.  

The list of parameter sets consists of (in order of publication) SKRA \cite{Ras00}, KDE0v1 \cite{Agr05}, NRAPR \cite{Ste05}, LNS \cite{Cao06}, and SQMC700 \cite{Gui06}.  Along with a short discussion of the different sets in this section, we comment on the various different subsidiary parameters and choices that occur when using Skyrme forces in finite nuclei.  In particular, one must choose how to include the exchange part of the Coulomb force, which is sometimes either ignored or treated in the Slater approximation (or rarely, more exactly), whether the contribution to the spin-orbit forces from the $t_1$ and $t_2$ terms is included and how the center of mass is treated.  Where these choices are not clear from the original papers, we have attempted to deduce them and commented as appropriate.  Where the papers are explicit, we do not repeat the information here.

The premise of SKRA\cite{Ras00} was that its equation of state in nuclear matter should be fitted to that derived from a realistic potential, along with relativistic and many-body corrections.  In this way, it was argued, the effective nucleon in the mean-field should contain as correlated a version of the bare nucleon as possible.  Ground state properties of doubly-magic nuclei were also included in the parameter fit.  This parameterization has been little explored in the literature since its exposition.  For SKRA, along with the fits to infinite nuclear matter, the finite nuclear calculations presented  were explicitly indicated as being performed with a particular published code \cite{hafom}.  From this it can be deduced that the finite nuclear calculations included the Coulomb exchange in the Slater approximation, included the $t_1$ and $t_2$ contributions to the spin-orbit force and used the diagonal approximation to the center of mass correction.

The KDE0v1 \cite{Agr05} parameterization had two underlying emphases in its derivation;  that the simulated annealing technique was used in optimizing the parameters, and that a wide range of observables from finite nuclei with a small amount of nuclear matter information were used in the parameter fitting.  The fitting observables include ground state energies and radii, as well as breathing mode energies.   Interestingly, other parameter sets appeared in the same paper as KDE0v1, which differed only in their starting parameters in the fit algorithm, but did not pass the nuclear matter constraints \cite{Dut12}.  This speaks of the non-linear nature of the fitness functions in terms of the parameters.  The KDE0v1 parameterization has a particular prescription for the center of mass correction, dependent on identifying the spherical quantum numbers of each occupied level.  We have not implemented this correction in the presented results, but in the spirit of KDE0v1, use an a posteriori treatment, taking the diagonal part of the center of mass operator.

NRAPR \cite{Ste05} aims, like SKRA, to fit an equation of state derived from realistic interactions, including pure neutron as well as symmetric nuclear matter, along with some adjustment to finite nuclei. 

LNS \cite{Cao06} again is fitted to nuclear matter calculations with realistic interactions - in this case to the equation of state and effective mass arising from Bruckner-Hartree-Fock calculations.  Adjustment is made taking a small set of doubly-magic finite nuclei into account.  Whether the $t_1$ and $t_2$ spin-orbit term is taken into account is not made explicit, and we present some results with both choices made.

SQMC700 \cite{Gui06}, which also makes the shortlist is based on a quark-meson coupling model which can be recast in the form of a density functional with certain assumptions.  The form of the functional is close to, but not exactly the same as the Skyrme functional.  Consequently, this force has so far been omitted from our calculations, though we intend to study it for a follow-up paper.  We note that the authors presenting the SQMC700 force surveyed properties of finite nuclei \cite{Gui06} in their original exposition.

The spin-orbit part of the force in all cases is fitted to finite nuclei.  Usually this is one or more single-particle spin-orbit splittings as deduced from the energy levels of odd-mass nuclei neighboring the doubly-magic nuclei, though NRAPR adjusts the spin-orbit strength to better reproduce binding energies of finite nuclei, without mentioning spin-orbit splittings.

\begin{figure}[tbh]
\begin{tabular}{cc}
  \includegraphics[width=.5\textwidth]{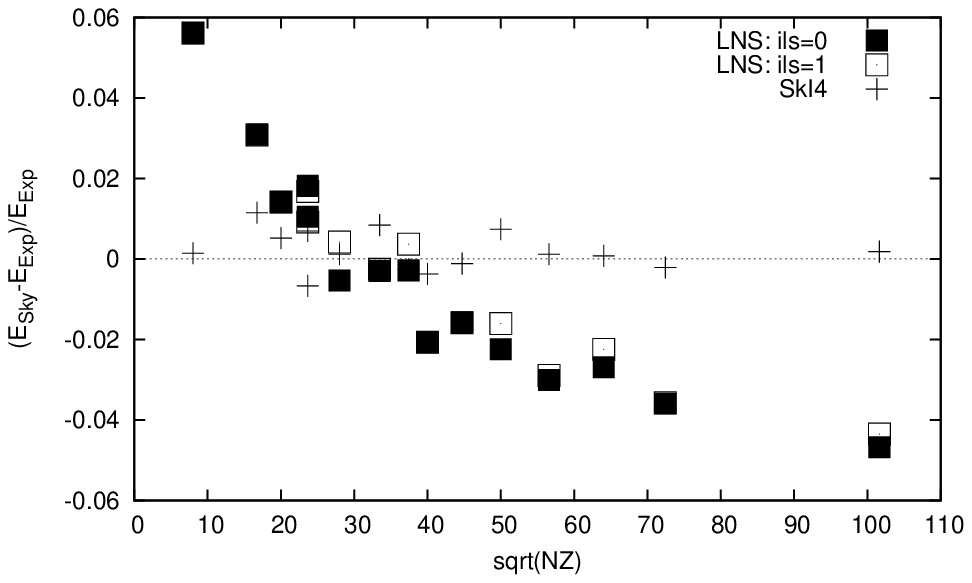} & \includegraphics[width=.5\textwidth]{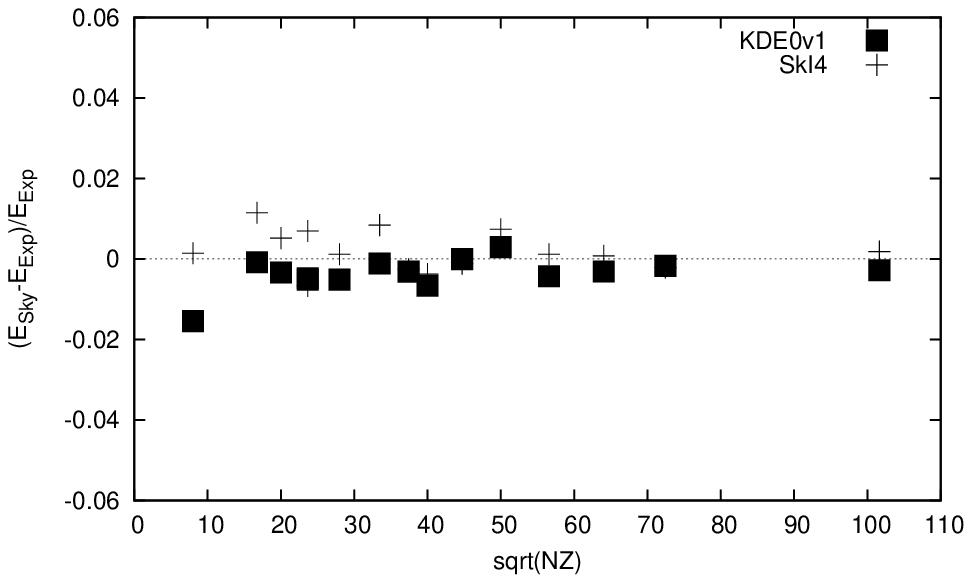}
\\
  \includegraphics[width=.5\textwidth]{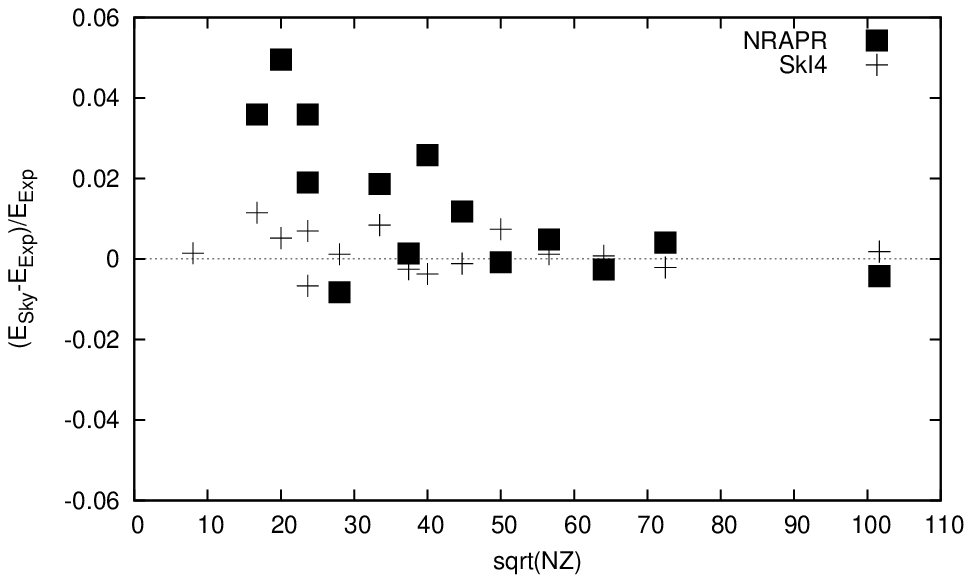} & \includegraphics[width=.5\textwidth]{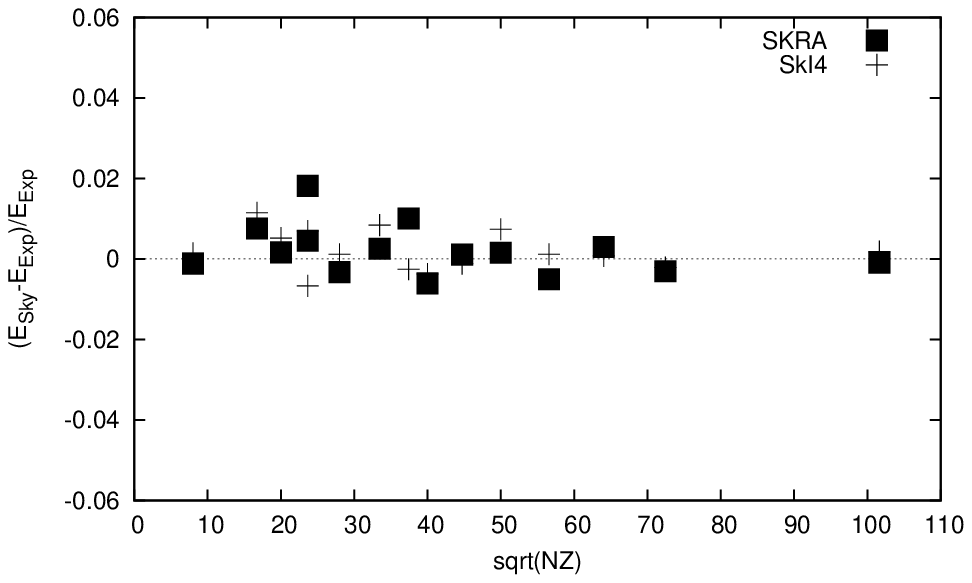}
\end{tabular}
  \caption{Binding energies of doubly-(semi)-magic nuclei for four shortlisted forces, compared to SkI4 as a reference.  The forces are labeled in the legend and described in the text.  For LNS (top left), ``ils=0'' means that the $t_1$ and $t_2$ term is not included in the spin-orbit force, ``ils=1'' means that it is included.}\label{fig:be}
\end{figure}

\section{observables}
We evaluate the following list of observables, motivated  by the selection in a recent review \cite{stone-reinhard}.  These are taken as a sample of a longer list of observables as a proxy to measure the success of forces which fit nuclear matter to reproduce finite nuclear properties.
\begin{enumerate}
\item Binding energies of even-even doubly-(semi)-magic nuclei.

since not all forces are fitted to these nuclei.  These are compared with a reference force, SkI4, which gives decent results.  Nuclei chosen are: $^{16}$O, $^{34}$Si, $^{40}$Ca, $^{48}$Ca, $^{48}$Ni, $^{56}$Ni, $^{68}$Ni, $^{78}$Ni, $^{80}$Zr, $^{90}$Zr, $^{100}$Sn, $^{114}$Sn, $^{132}$Sn, $^{146}$Gd and $^{208}$Pb.  In plots, the independent variable is given as $\sqrt{NZ}$ to help separate the points.

\item Fission barriers.

Early Skyrme forces had not been particularly successful in reproducing the fission barriers in heavy nuclei, prompting the widely-used SkM* parameterization, fitted to give a good description of the deformation path of $^{240}$Pu \cite{skms}. We apply the same pairing prescription to each force, being a volume delta pairing with an averaged strength \cite{bender-review}.  We compare with the SkM* force as a reference, as it was fitted with the fission path in mind.

\begin{figure}[tbh]
  \includegraphics[width=.8\textwidth]{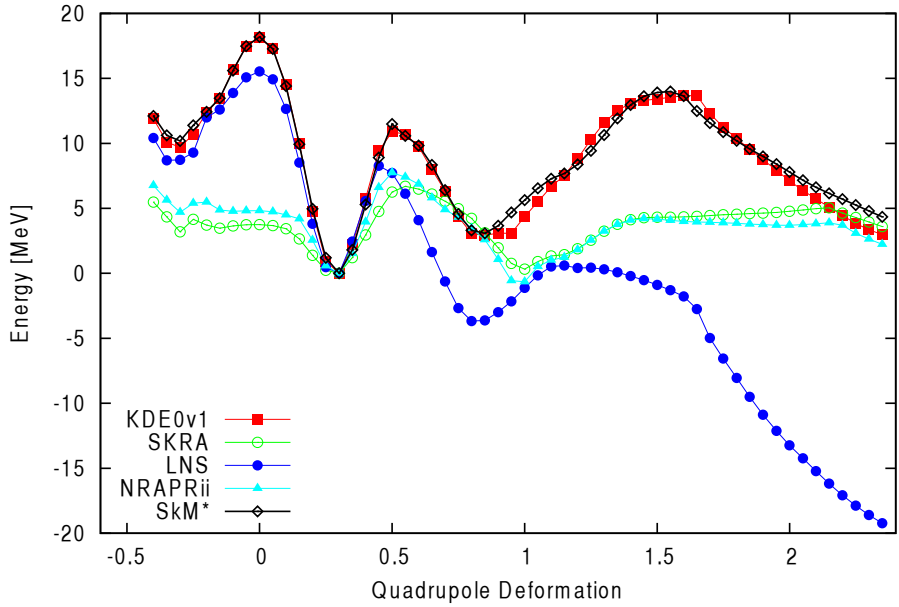}
  \caption{Fission barrier in $^{240}$Pu for the subset of forces, compared with SkM*\label{fig:fb}}
\end{figure}

\item Isotope shifts

The changes in radii across an isotopic chain is something which is very much in the realm of finite-nuclear properties, and dependent on the details of the single-particle energies, so it is interesting to see the extent to which forces fitted mostly to nuclear matter data can perform well in such observables.  We choose a particularly difficult example, namely the isotope shift between the two doubly-magic calcium nuclei $^{40}$Ca and $^{48}$Ca.  The chain of even-even isotopes between these two nuclei show a roughly parabolic behavior in the radii, with $^{40}$Ca and $^{48}$Ca having similar charge radii ($r^2$($^{48}$Ca)$-r^2$($^{40}$Ca)=0.007 fm$^2$ \cite{ski4}), with an increase in between as the $f_{7/2}$ shell is partially filled.  This general trend is poorly described by Skyrme forces, which is usually blamed on correlation effects.  Other methods have been shown to reproduce the trend in certain circumstances \cite{caurier,fayans}.  We do not check the whole trend here, just the shift between the two specific doubly-magic isotopes.
\end{enumerate}

\section{Results}

\subsection{Binding energies}
The binding energies are shown in Figure \ref{fig:be}.  As a reference, results from SkI4 \cite{ski4}, as an example of a force fitted to ground-state data of many doubly-magic nuclei.  A range of results is evident, from LNS, with a somewhat poor isospin-dependence, through NRAPR, which has a spread larger than the reference SkI4, to SKRA and KDE0v1 which are very competitive with the reference force.  Curiously, the small number of forces which pass the strict nuclear matter constraints, show quite different behavior for binding energies of finite nuclei.  

\subsection{Fission Barriers}
Figure \ref{fig:fb} shows the deformation energies for axially-deformed constrained Hartree-Fock calculations for the subset of four forces, with SkM* as a reference.  We use a slightly modified NRAPR force called NRAPRii, in which the spin-orbit parameter has been doubled.  The standard NRAPR has a spin-orbit force so weak that even $^{208}$Pb is not doubly-magic.  Doubling the strength fixes this without breaking the nuclear matter tests.  The forces group into three sets;  LNS, which had the poorest isospin dependence in binding energies does not reproduce the fission barrier structure at all.  It has the hyper-deformed minimum as being the ground state, with no further barrier to the fission path.  SKRA and NRAPRii look very similar, but with weak barriers, while KDE0v1 reproduces the SkM* very well.  This is all the more remarkable given that SkM* was designed to account for the fact that standard Skyrme forces, fitted to ground state data, did not reproduce this barrier.

\begin{figure}[tbh]
  \includegraphics[width=.8\textwidth]{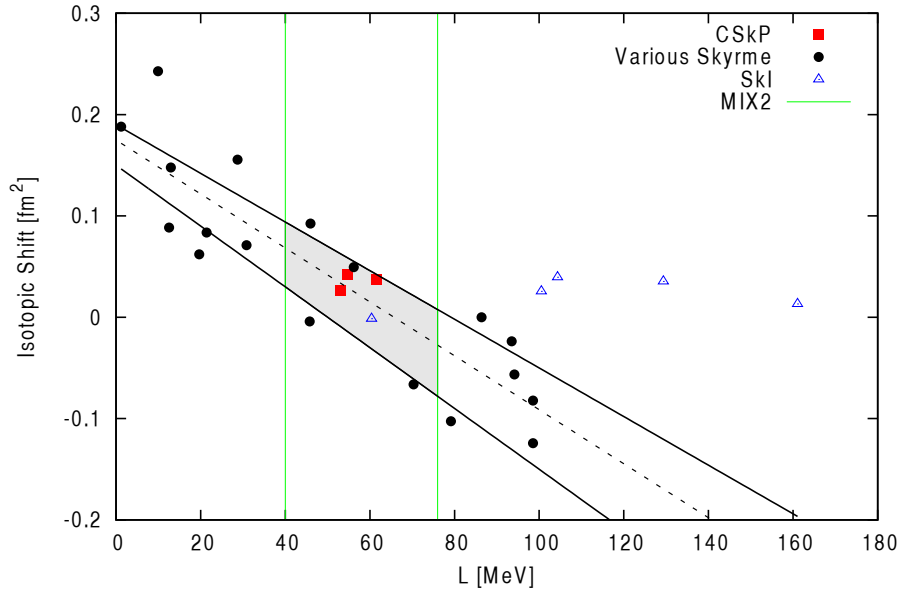} 
  \caption{Isotope shift between $^{40}$Ca and $^{48}$Ca for various Skyrme forces, including the subset, labeled as {\it CSkP}, various other Skyrme forces and in particular the SkI forces, which were fitted to this observable.\label{fig:is}}
\end{figure}

\subsection{Isotope Shifts}
The Calcium isotope shifts are shown in Figure \ref{fig:is}.  They are plotted against the derivative, $L$, of the symmetry energy at saturation density, since this has been shown to correlate with radius data in lead isotopes \cite{chen}.  The dashed line in the plot shows the line of best fit, with the solid lines indicating one standard deviation in the fit.  The vertical lines, labeled {\it MIX2} show the range allowed in the nuclear matter constraints \cite{Dut12}.  Obviously the subset of forces which passed the constraints appear in the allowed region.  The correlation between $L$ and the isotope shift is quite strong, as previously shown in lead, and gratifyingly the previously allowed values of $L$ are consistent with the observed isotope shift in calcium.  The outliers in the calculation are the SkI forces \cite{ski4}, which were fitted to the isotope shift, but not to $L$.  Although it appears that Skyrme forces will generally correlated $L$ with the isotope shift, one can bypass this correlation if one chooses, by including suitable observables in the fit.  It would be interesting to repeat this exercise for the lead isotopes to see if the correlation is easily broken by suitable fitting.

\section{Conclusions}
A subset of Skyrme forces, which passed a series of tests on nuclear matter properties derived from data, have been analyzed for their behavior in finite nuclei.  It is found that for observables with no direct correlation to nuclear matter quantities, a range of different behavior is found, while for a particular example of a finite nuclear observable with a known correlation to nuclear matter properties, passing the constraint seems sufficient, but not necessary, to fit the experimental data.  The KDE0v1 force fits the finite nuclear data presented quite well, and it was also the only force presented fitted to a signigicant amount of data from finite nuclei.  It was also the only force able to pass the nuclear matter constraints of \cite{Dut12} without inclusion of a wide range of Equation of State data in the fit.  Tentatively, one can conclude that Skyrme parameterizations cannot be necessarily expected to reproduce data, to which they were not fitted, with very high accuracy.  

\begin{theacknowledgments}
Financial support from the UK STFC, and from the Santander University Network is acknowledged in support of this work.  MD thanks the Brazilian Agency FAPESP.
\end{theacknowledgments}



\bibliographystyle{aipproc}   


\IfFileExists{\jobname.bbl}{}
 {\typeout{}
  \typeout{******************************************}
  \typeout{** Please run "bibtex \jobname" to obtain}
  \typeout{** the bibliography and then re-run LaTeX}
  \typeout{** twice to fix the references!}
  \typeout{******************************************}
  \typeout{}
 }



\end{document}